\documentclass[times,review]{elsarticle}

\usepackage{framed,multirow}
\usepackage{mhchem}
\usepackage{amsmath}

\usepackage{relsize} 

\usepackage[final]{changes}

\definechangesauthor[name={Referee 1}, color=orange]{R1}
\definechangesauthor[name={Referee 2}, color=blue]{R2}
\colorlet{Changes@Color}{red}

\usepackage{todonotes}
\makeatletter
\setremarkmarkup{\todo[color=Changes@Color#1!20,size=\scriptsize]{ #2}}
\makeatother

\usepackage{amssymb}

\DeclareMathOperator\erf{erf}

\usepackage{latexsym}
\usepackage{mathtools}
\usepackage[cdot,squaren,Gray,binary,notextstyle]{hepunits} 
\usepackage{sistyle}
  \SIunitspace{\cdot}
  \SIunitdot{\cdot}
  \unitskip{\cdot}

\usepackage{subcaption}
\usepackage[labelsep=quad,skip=3pt]{caption}
\usepackage{color}

\usepackage[ruled,linesnumbered]{algorithm2e}
\SetKwRepeat{Do}{do}{while}
\DontPrintSemicolon

\definechangesauthor[name={xx},color=red]{xx}

\journal{Journal of Computational Physics}

\begin{document}

\begin{frontmatter}

\title{Rejection-based sampling of inelastic neutron scattering}%

\author[1,2]{X.-X. Cai\corref{cor1}}
\ead{xcai@dtu.dk}
\cortext[cor1]{Corresponding author:}

\author[2]{T. Kittelmann}
\author[1,2]{E. Klinkby}
\author[3]{J.I. M\'{a}rquez Dami\'{a}n}

\address[1]{Technical University of Denmark, Denmark}
\address[2]{European Spallation Source ERIC, Sweden}
\address[3]{Nuclear Data Group, Neutron Physics Department, Centro At\'{o}mico Bariloche, CNEA, Argentina}


\begin{abstract}
Distributions of inelastically scattered neutrons can be  quantum dynamically described by a scattering kernel. We present an accurate and computationally efficient rejection method for sampling a given scattering kernel of any isotropic material. The proposed method produces continuous neutron energy and angular distributions, typically using just a single interpolation per sampling. We benchmark the results of this method against those from accurate analytical models and one of the major neutron transport codes. We also show the results of applying this method to the conventional discrete double differential cross sections.

\end{abstract}


\end{frontmatter}


 
\section{Introduction}

The \deleted{neutron} scattering kernel~\cite{hove1954}, also known as the scattering law or the \added{inelastic part of the} scattering function, is defined by the  dynamical properties of a material \deleted{and the associated scattering lengths of the probing neutron} and is a function of energy- and momentum-transfers. 
Described by a scattering kernel, the vibrational excitations and atomic diffusive motions are fingerprinted in the double differential scattering cross sections. 
Concerning neutron scattering in condensed matter, scattering kinematics are often simulated by Monte Carlo methods to relate measurements to underlying physics. 
When the distributions related to individual scatterings are important, such as when performing data correction or characterisation at neutron scattering instruments~\cite{Farhi2009}, the scattering is sampled directly from the scattering kernel in order to reproduce the measured double differential cross sections. 
The energy region of interest in such applications is typically below 100 milli-electronvolt. 
On the other hand, in neutronics applications, e.g.,~\cite{MacFarlane2010}, scattered neutron states are traditional sampled from the pre-processed double differential cross sections in either continuous or discrete forms for incident neutrons up to a few electronvolt. 
However, in recent years, attracted by its small memory footprint and the absence of geometrical limits to its validity, the method of directly sampling from scattering kernels~\cite{bischoff1972} has been implemented in a few neutronic codes as well~\cite{Ballinger1999,Pavlou2014,hart2017}. 

The concept of rejection-based Monte Carlo sampling is well suited for neutron scattering, as the analytical expression
required for sampling via the transformation method
 generally does not exist. 
However, if the bounding distribution is not chosen with care, the computational efficiency of this method may be unacceptable. For example, a prior attempt of sampling 1eV neutron scattering with hydrogen in water showed less than 0.1\% acceptance rate~\cite{Ballinger1999}.

The scope of this paper is to present an accurate and efficient method for sampling a given scattering kernel.
Section~\ref{sMethod} introduces our method to sample the kernels for isotropic materials. 
For the purpose of verifying the rejection method presented here, we compare its predictions with results from analytical models and one of the major Monte Carlo neutron transport codes in section~\ref{sVerification}.
Methods of applying this rejection method to the commonly used discrete cross sections is also discussed in this section. 
We conclude this work in section~\ref{sConclusion}.

\section{Method}
\label{sMethod}

Given \replaced{an isotropic}{ a} scattering kernel $S$, for a material at temperature T, the relationship between bound and double-differential cross section per atom is given by~\cite{bischoff1972}~\footnote{For simplicity, $S$ used in this work is assumed to be an asymmetric scattering function, but it would be straight-forward to apply the presented sampling method to symmetric scattering kernels.}
\begin{equation}
\label{eDoubleDiff}
\frac{ \mathrm d ^2 \sigma}{ \mathrm d  E^\prime  \mathrm d  \Omega}
\deleted{=\frac{\sigma_b}{4\pi} {\frac{k^\prime}{k}} S(Q, \omega)}
= \sqrt{\frac{E^\prime}{E}} \frac{\sigma_b}{4\pi} \frac{S(\alpha, \beta)}{ k_bT}
\end{equation}
with 
\begin{equation}
\label{eAB}
\alpha =\frac{E+E^\prime-2\mu\sqrt{E^\prime E}}{k_bT}  \quad\text{and}\quad  \beta= \frac{E^\prime-E}{k_bT} 
\end{equation}

Here $\alpha$ is the reduced momentum transfer, $\beta$ is the reduced energy transfer, $\sigma$ and $\sigma_b$ are the scattering and bound cross sections, respectively, $\mu$ is the cosine of the scattered angle, $k_b$ is the Boltzmann constant, $T$ is the temperature, $E$ and $E^\prime$ are the incident and scattered energies, respectively\deleted{, and $k$ and $k^\prime$ are the wavevectors before and after the scattering}. 
Notice that the definition of $\alpha$ here differs from that in NJOY~\cite{MacFarlane2010}, which mainly concerns incoherent scattering kernels, by the factor $M/m$, where $M$ and $m$ are the target and neutron masses, respectively. 
Inclusion of target masses in the formulas is not problematic for incoherent scattering, where separate contributions from scattering on atoms with
different masses is additive. But in a coherent kernel, scattering
originates from correlations of different atoms, and the definition of a
target mass is thus ambigous for any material which does not consist of a
single isotope.
In either case, however, to represent a valid scattering, $\alpha$ and $\beta$ should satisfy two conditions: 
firstly, the scattered energy must be non-negative, thus the value of $\beta$ should not be less than \replaced{$\beta_-=-E/k_bT$}{$\beta_-=-E/kT$}; secondly, the cosine of the scattered angle must be in $[-1,1]$.  Accordingly, the lower and upper bounds on $\alpha$ can be expressed as a function of energy transfer~\cite{bischoff1972}.
\begin{equation}
\label{eAlimts}
\alpha_\pm (\beta) = \frac{2E + k_bT\beta \pm 2\sqrt{E(E+k_bT\beta)}  }{k_bT}
\end{equation}

The integral scattering cross section can be evaluated as~\cite{bischoff1972}
\begin{equation}
\label{eTotalCross}
\sigma(E)= \frac{C}{E} \int\displaylimits_{-\infty}^\infty \!\int\displaylimits_{-\infty}^\infty     S(\alpha, \beta) \; \Theta (\alpha, \beta) \, \mathrm d  \alpha \, \mathrm d \beta 
\end{equation}
where $C$ is a constant that equals ${\sigma_b k_b T}/{4}$, and \added{$\Theta$ is the mask for the scattering phase-space in a general material.}
\begin{equation}
\label{eThta}
\Theta (\alpha, \beta) =\begin{cases}
           \, 1  &\text{when} \;\; (\alpha, \beta) \in D \\
           \, 0  &   \text{otherwise}
            \end{cases}
\end{equation}
Here the region $D$ \replaced{confines }{represents} the entire valid scattering phase-space of incident neutrons at $E$, and can be expressed as
\begin{equation}
\label{eCut}
D=  \left\{ (\alpha, \beta) : \; \beta \ge -E/k_bT \;\;\; \text{and}\;\;\; \alpha_-(\beta) \le  \alpha   \le \alpha_+(\beta) \; \right\}
\end{equation}
It can be shown that the signs of the first derivative of $\alpha_-$ and $\alpha_+$  are non-positive and positive, respectively, for any incident energy, while  $\beta_-$ obviously decreases with increasing incident energy. \replaced[id=R1, remark=R1C3]{Therefore, $D$ is a monotonically expanding region with increasing incident energy. 
On the other hand, a scattering kernel $S$ is defined only by the material properties, and those are not dependent on the incident particle nor the particle interaction potential with the material~\cite{hove1954}.
As a result, the volume of the double integral in Eq.~\ref{eTotalCross}, which is equal to $\sigma(E)E / C$, is also monotonically increasing with increasing incident energy. 
Notice, in practice, measurements may trade off the size covered by $D$ against better resolution or signal to noise ratio. Incident energy is often tuned according to the material properties under investigation and the particular instrument setup.
}{In other words, $D$ is a monotonically expanding region with increasing incident energy.}

In terms of sampling, it can be seen from Eq.~\ref{eTotalCross} that the conditional probability density function (PDF) is
\begin{equation}
\label{eKinematic}
p(\alpha,\beta \, | E) = \frac{C}{\sigma(E)E} S(\alpha, \beta) \, \Theta (\alpha, \beta) 
\end{equation}
It is relatively computational \replaced{expensive }{intensive} to construct such a distribution in either continuous or discrete forms, thus  on-the-fly determination of $p$ is often avoided in Monte Carlo simulations. 
Numerically,  this distribution has instead been approximated by interpolating the sampled results, from the distribution of a slightly higher energy neutron and that of a slightly lower energy neutron, with respect to the actual incident energy.  
As a better alternative, we here propose a sampling method that can skip this approximation, thus eliminating potential errors associated with the interpolations.  
To do so, it is desired to develop the PDF further into the form of
\begin{equation}
p(x)=A^{-1}p_1(x)\,h(x)
\end{equation}
where $A^{-1}$ is a normalisation factor, \replaced[id=R1,remark=R1C5]{$p_1$}{$p$} is a PDF of $x$ on a two dimensional space, and $h$ is a scalar function in the range between 0 and 1. 
Suggested by item R13 in~\cite{Everett1983},  a PDF in this form can be sampled by the rejection-based Monte Carlo method.
Briefly, the first step is to sample a value $x$ from $p_1$, along with a random number $\xi$ drawn uniformly from $[0,1\replaced{)}{]}$, then either accept $x$ if $\xi \leq h(x)$, or sample a new value of $x$ and repeat the procedure. The efficiency of this, i.e.\added{,} the acceptance rate, is $A$. 

Let $E_l$ and $E_h$ be the kinetic energies of two neutrons that satisfy $E_l < E_h$. Assuming the distribution for $E_h$ is given, and only the scattered status of a neutron at $E_l$ is of interest. From Eq.~\ref{eKinematic}, we have
\begin{equation}
\label{eRatio}
\frac{p(\alpha,\beta \, | E_l)}{p(\alpha,\beta \, | E_h) }=\frac{ \sigma(E_h)E_h S(\alpha, \beta) \, \Theta_l (\alpha, \beta) }{ \sigma(E_l)E_l S(\alpha, \beta) \, \Theta_h (\alpha, \beta) }
\end{equation}
Notice that Eq.~\ref{eRatio} is only well-defined within the region $D_h$. However, because of the energy-dependent expansion behaviour of $D$, $D_h$ already covers the full region of interest, i.e.\added{,} $D_l$,  within which valid samples are all situated. In the region $D_h$, $\Theta_h$ becomes unity, we can obtain
\begin{equation}
\label{eTheRealDist}
p(\alpha,\beta \, | E_l) = \frac{\sigma(E_h) E_h}{\sigma(E_l) E_l} p(\alpha,\beta \, | E_h) \, \Theta_l (\alpha, \beta) 
\end{equation}
To sample from Eq.~\ref{eTheRealDist}, we first sample a pair of $(\alpha,\beta)$ at $E_h$, i.e.\added{,} $p(\alpha,\beta \, | E_h)$, and then consider the acceptance of this sample according to $\Theta_l$. 
Unlike the $h$ function in the general case, its equivalent function, $\Theta$, in this case equals either 0 or 1.
Therefore, the generation of the random number $\xi$ can be skipped. 
The acceptance rate in the case of Eq.~\ref{eTheRealDist} is
\begin{equation}
\label{eEfficientcy}
\mathcal{M}(E_l,E_h) = \sigma(E_l)E_l/\sigma(E_h)E_h
\end{equation}
For a special case, in the so-called $1/v$ region, where the cross section is inversely proportional to the neutron incident speed, 
$\mathcal{M}$ is simply $\sqrt{E_l/E_h}$.
When a dense energy grid in the evaluated data library is used,  $\mathcal{M}$ is generally greater than 90\% in the thermal neutron range.

In the rejection method, it is clear that the probability of accepting a sample for a neutron at a certain energy is $\mathcal{M}$. 
In terms of computational speed, the cost on the samples that are eventually dropped can be less than that, as sampling algorithms can be optimised by rejecting unsatisfied samples at an early stage, before the expensive part of the sampling algorithm. By default, all simulated data in the later sections of this paper are generated using Algorithm~\ref{AFast}. 
However, a shortcoming of Algorithm~\ref{AFast} is that the timing behaviour is less straight forward to predict. 
As the worst case scenario for speed, the Algorithm~\ref{APredictable}, where the computational speed is proportional with the acceptance rate, is also implemented to help verifying the timing behaviour of the rejection method. 

\noindent\makebox[\textwidth][c]{%
\begin{minipage}[t]{0.45\textwidth}
  \vspace{0pt}
  \begin{algorithm}[H]
    \label{AFast}  
    \caption{optimised for speed}
        find smallest i, so that $E \leq E_i$ \\
        \Do{$\alpha^\prime \notin\left[\alpha_-,\alpha_+\right]$}{                     
        \Repeat{$-E/k_bT < \beta ^\prime  $}{
             sample a $\beta^\prime$ from $P(\beta\,|\,E_i)$\\
            }
    find j, so that $\beta_{j-1} \leq \beta^\prime \leq \beta_j$ \\
    sample $\alpha_l$ from $F(\alpha \,|\, \beta_{j-1},E_i)$\\
    sample $\alpha_h$ from $F(\alpha \,|\, \beta_{j},E_i)$\\
    interpolate an $\alpha^\prime$ linearly from  $\alpha_l$ and  $\alpha_h$\\
        }
        accept $\alpha^\prime$ and $\beta^\prime$  \\
  \end{algorithm}
\end{minipage}
\hspace{0.05\textwidth}
\begin{minipage}[t]{0.45\textwidth}
  \vspace{0pt}  
  \begin{algorithm}[H]
    \label{APredictable}
    \caption{predictable speed}
    find smallest i, so that $E \leq E_i$ \\
     \Do{$\alpha^\prime \notin\left[\alpha_-,\alpha_+\right]$ \textrm{or} ${-E}/{k_bT}  \geq \beta ^\prime $}
     {
    sample a $\beta^\prime$ from $P(\beta\,|\,E_i)$\\
    find j, so that $\beta_{j-1} \leq \beta^\prime \leq \beta_j$ \\
    sample $\alpha_l$ from $F(\alpha \,|\, \beta_{j-1},E_i)$\\
    sample $\alpha_h$ from $F(\alpha \,|\, \beta_{j},E_i)$\\
    interpolate an $\alpha^\prime$ linearly from  $\alpha_l$ and  $\alpha_h$\\
     }
    accept $\alpha^\prime$ and $\beta^\prime$ 
  \end{algorithm}
\end{minipage}%
}

\subsection{Numerical approximations employed in the implementation}
\label{ssNum}

To evaluate the double integral in Eq.~\ref{eTotalCross}, we make a few approximations, which are independent from the proposed sampling method. 

\added[id=R2, remark=R2 C6]{Denoting the distribution of $\alpha$ at a given $\beta$ as $S(\alpha \, | \, \beta)$}, we define $\mathcal{F}$ as follows~\cite{bischoff1972} 
\begin{equation} 
\label{eFFunc}
\mathcal{F}(\alpha_x \, | \, \beta) = \int_{0}^{\alpha_x} S(\alpha \, | \, \beta)  \mathrm d \alpha
\end{equation}
As a kernel is tabulated in a table in practice, $\mathcal{F}(\alpha_x \, | \, \beta)$  consists of  piecewise functions that are continuous in the $\alpha$ space. 
At the $n_{th}$ point of an $\alpha$ grid, $\mathcal{F}$ can be evaluated as
\begin{equation} 
\label{eFFuncReal}
 \mathcal{F}(\alpha_n \, | \, \beta) = \sum_{k=0}^{n-1}\int_{\alpha_{k}}^{\alpha_{k+1}} S(\alpha \, | \, \beta)  \mathrm d \alpha
\end{equation}
To be compatible with the ENDF standard, it is assumed that $S(\alpha \, | \, \beta)$ can be interpolated by the log-linear law~\cite{MacFarlane2010,trkov2012endf}, using which gives the following evaluation for the $\alpha$ integral over an interval $[\alpha_1, \alpha_2]$  
\begin{equation}
\label{eAlphaIntegral}
\int\displaylimits^{\alpha_2}_{\alpha_1} \, S(\alpha \, | \, \beta) \mathrm{d} \alpha =  \, \frac{(\alpha_2-\alpha_1)[S(\alpha_2 \, | \,\beta)-S(\alpha_1\, | \,\beta)]}{\mathrm{log}[S(\alpha_2\, | \,\beta)/S(\alpha_1\, | \,\beta)]}
\end{equation}
In the case where Eq.~\ref{eAlphaIntegral} is not well-defined, the integral is evaluated as $(\alpha_2-\alpha_1)[S(\alpha_2 \, | \,\beta)+S(\alpha_1\, | \,\beta)]/2$ instead. 

The cumulative distribution of $\alpha$ can be expressed as
\begin{equation}
\label{eCumuAlpha}
 F(\alpha \, | \, \beta, E) = \frac{\mathcal{F}(\alpha \, | \, \beta) - \mathcal{F}(\alpha_- \, | \, \beta)}{\mathcal{F} (\alpha_+ \, | \, \beta) - \mathcal{F} (\alpha_- \, | \, \beta)} 
 \quad\text{with}\quad  \alpha\in[\alpha_-, \alpha_+]
\end{equation}

To sample an $\alpha$, we  generate a uniform random number $\xi_1$ from $[0,1)$, and solve for $\alpha$ according to $\xi_1=F(\alpha \, | \, \beta, E)$.
The first step is to find the  greatest index $k$ satisfying $\xi_1 > F(\alpha_k \, | \, \beta, E)$, so the distribution of the pursued $\alpha$ is  the piecewise function confined in $[\alpha_k, \alpha_{k+1}]$. 
The cumulative density function of $\alpha$ in the piecewise function can be calculated as
\begin{equation}
\label{eFindPiece}
r=  \frac{\xi_1 -F(\alpha_k \, | \, \beta, E) }{F(\alpha_{k+1} \, | \, \beta, E) - F(\alpha_{k} \, | \, \beta, E)}
\end{equation}

If Eq.~\ref{eAlphaIntegral} is not well-defined for the piecewise function in $[\alpha_k, \alpha_{k+1}]$, the distribution is treated as a linear function. 
Otherwise, inverting Eq.~\ref{eAlphaIntegral} to solve for $\alpha$ leads to  
\begin{equation}
\label{eSampling}
\alpha = -x y^{-1} \log \Big[ \frac{
S(\alpha_k \, | \, \beta) \exp (-\alpha_k y/x)
}{r \,[(S(\alpha_{k+1} \, | \, \beta)-S(\alpha_k \, | \, \beta) ]+ S(\alpha_k \, | \, \beta)} \Big] \quad\text{with}\quad x=\alpha_{k+1}-\alpha_k \quad\text{and}\quad  y=\log \Big[ \frac{S(\alpha_{k+1} \, | \, \beta)}{S(\alpha_k \, | \, \beta)} \Big]
\end{equation}
In the implementations of both Algorithm~\ref{AFast} and~\ref{APredictable}, $y$ is pre-computed and cached in memory for the entire $(\alpha, \beta)$ grid.

For the outer $\beta$ integral of Eq.~\ref{eTotalCross}, a simple trapezoidal approximation is used over all $\beta$ intervals. The integral in $[\beta_1,\beta_2]$ is evaluated as
\begin{equation}
\label{eTrap}
\int\displaylimits_{\beta_1}^{\beta_2}  \!\int\displaylimits_{-\infty}^\infty     S(\alpha, \beta) \; \Theta (\alpha, \beta) \, \mathrm d  \alpha \, \mathrm d \beta 
\approx 
\frac{1}{2}(\beta_2-\beta_1)
\left\{   \left[  \mathcal{F}\left(\alpha_+ \, | \, \beta_2\right) - \mathcal{F}\left(\alpha_- \, | \, \beta_2\right)  \right] +  \left[  \mathcal{F}\left(\alpha_+ \, | \, \beta_1\right) - \mathcal{F}\left(\alpha_- \, | \, \beta_1\right)  \right] \right\}
\end{equation}

It can be seen from Eq.~\ref{eTrap} that the inner integral is proportional with the probability density of $\beta$, so it in fact can be used to generate point-wise linear functions, which are useful for sampling $\beta$.
However, this low order trapezoidal approximation is often a major, if not the dominant, error source when approximating fine structures of energy transfer distributions, e.g., the quasi-elastic peak~\cite{Damian2014} in liquids that represents the diffusive motions.
To minimize the numerical errors, the $\beta$ grid used in caching is finer than the input kernel in some critical regions. 
\added[id=R2, remark=R2 C2]{The general procedure of such refinement is to increase the grid gradually and check for the convergent criteria, which depends strongly on the physics and applications.
As the kernels used in this work are given to be purely incoherent by nature or by applied approximation, the scattered momentum distributions are expected to be smooth.
We test the convergence of total cross section as the measure of $\beta$ grid refinement. 
The increased points in the kernel are interpolated in the log-linear law suggested by the ENDF standard.
Such treatment is expected to be adequate for the transport problems considered in this work. 
}

\section{Benchmark and verification}
\label{sVerification}

\subsection{Benchmarking the accuracy of implementation}

For the purpose of benchmarking our implementation, we study the scattering in a classical monoisotopic ideal gas, where the scattering kernel, the total cross section and the scattered energy distribution can be expressed analytically at closed-form expression.

The gas system of interest consists of motion independent nuclei in an ensemble at 293.6K. 
Each nuclei moves freely with the speed distribution of a Maxwellian. 
The scattering kernel can be expressed analytically~\cite{squiresnew}. 
In dimensionless form~\cite{MacFarlane2010}, it can be expressed as.
\begin{equation}
\label{eFreegasS}
S(\alpha,\beta) = \frac{\exp(-\beta/2)}{\sqrt{4\pi\alpha/A}} 
\exp\Bigg[{-\frac{\alpha^2/A^2 + \beta^2}{4\alpha/A}}\Bigg]
\end{equation}
where $A$ is the ratio between the nucleus mass and the neutron mass. 

For simplicity, if we assume the atomic mass is identical to the neutron mass, and the free scattering cross section is unity, the expressions of the total cross sections and energy differential cross section are particularly simple~\cite{Weston2007}. They are shown in Eq.~\ref{eXS} and \ref{eDiff}.

\begin{equation}
\label{eXS}
\sigma(E) =  (1+\frac{1}{2a^2}) \,\erf(a) + \exp(-a^2) / (a\sqrt{\pi}) \quad\text{with}\quad  a = \sqrt{E/k_bT}
\end{equation}

\begin{equation}
\label{eDiff}
  f(E \rightarrow  E^\prime)=\begin{cases}
     \exp(\beta) \erf(\sqrt{ E/\replaced{k_b}{k}t}) \, / [E\sigma(E)] &,\,\, E<E^\prime\\
             \erf(\sqrt{ E^\prime /\replaced{k_b}{k}t})   / [E\sigma(E)] &,\,\,E>E^\prime
            \end{cases}
\end{equation}

\begin{figure}
\centering
  \begin{subfigure}[b]{0.45\textwidth}
  \centering\includegraphics[height=0.25\textheight]{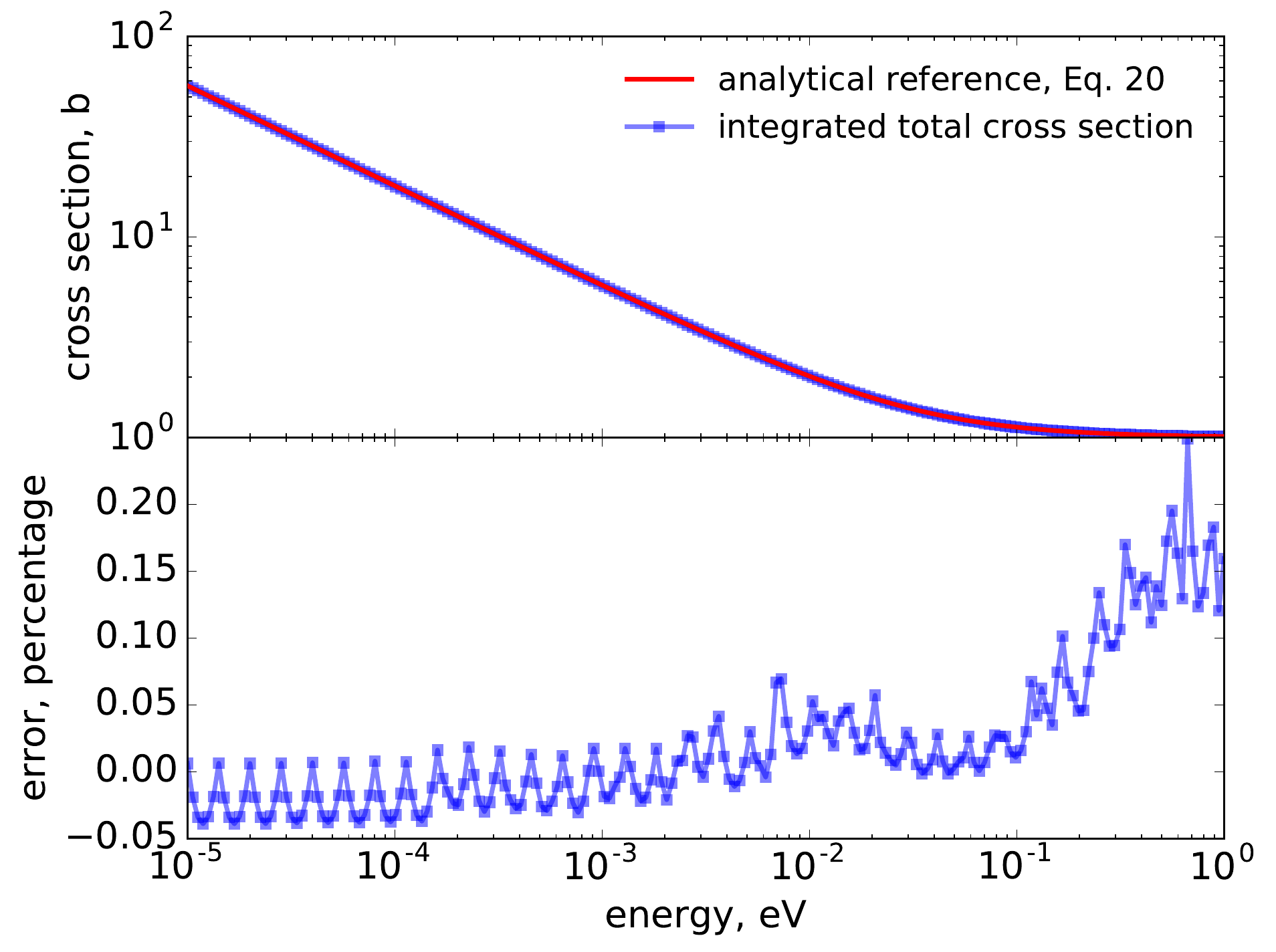}
    \caption{total cross section}\label{ffabs_error}
  \end{subfigure}
    ~~~~~
  \begin{subfigure}[b]{0.45\textwidth}
    \centering\includegraphics[height=0.25\textheight]{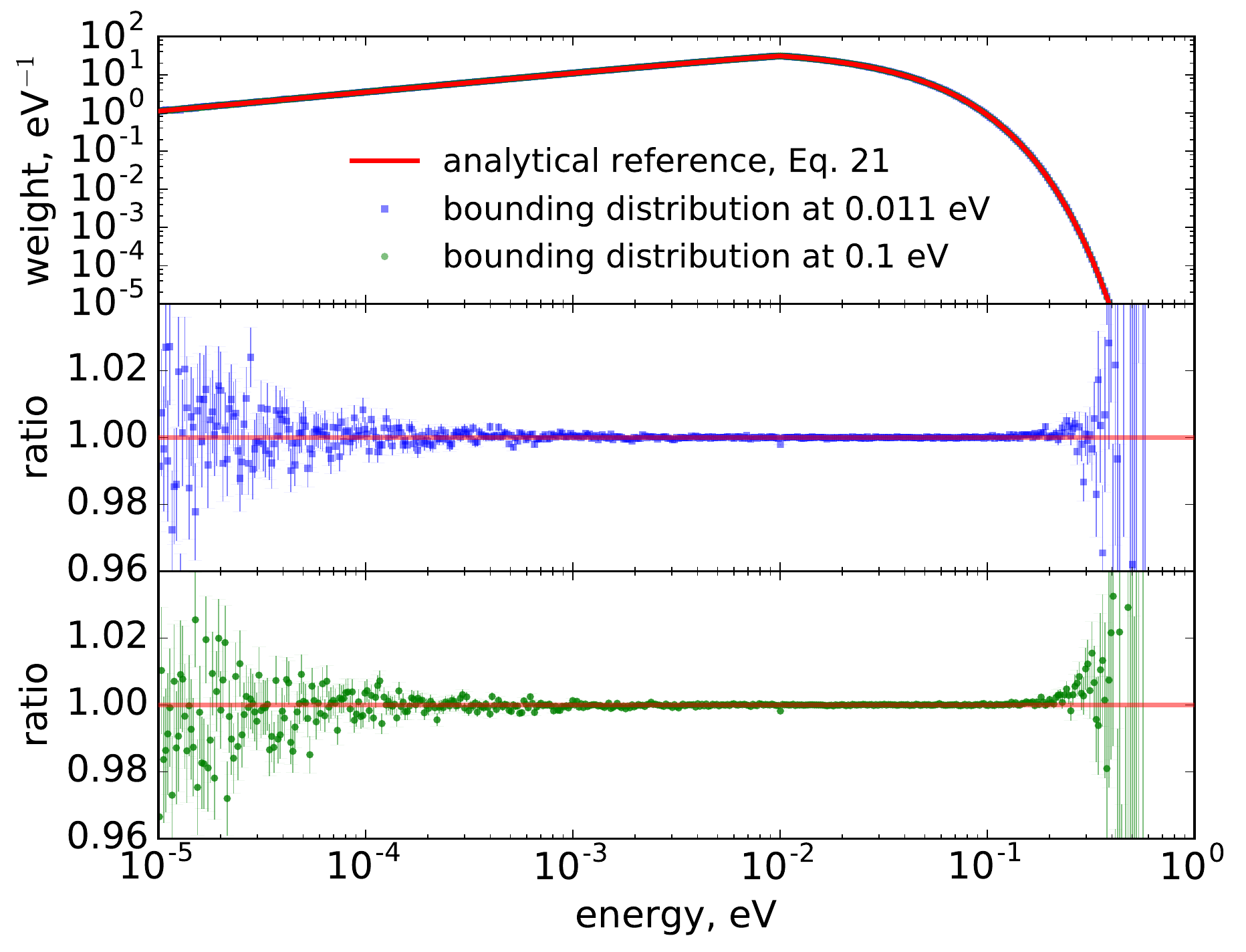}
    \caption{energy distribution}\label{ffdiff}
  \end{subfigure}
 \caption{Neutron scattering with the hypothetical ideal gas in three different setups.
 Fig.~\ref{ffabs_error} compares our implemented Eq.~\ref{eTotalCross} and the analytical reference Eq.~\ref{eXS}. 
 Fig.~\ref{ffdiff} shows the scattered energy distributions of ten billion  neutrons at \SI{0.01}{\eV} using the bounding distributions constructed at two higher energies.  Error bars represent the root-mean-squared error. Results are benchmarked against the analytical reference Eq.~\ref{eDiff}. 
 }
 \label{fMean}
\end{figure}

We simulate this hypothetical ideal gas and compare the results with these equations. 
The alpha and beta grids of the kernel used in the simulations are the same as those in the ENDF/B-VIII.0 H(H2O) TSL evaluation~\cite{Brown2018}, except for the density of the beta grid which is refined by a factor of two. 
The total cross section calculated by our implementation are compared with the analytical reference in Fig.~\ref{ffabs_error}. The absolute error of our implementation is smaller than 0.05\% below about \SI{0.1}{\eV}, trending towards 0.2\% at \SI{1}{\eV}. 
Fig.~\ref{ffdiff} benchmarks the scattered energy distribution of \SI{0.01}{\eV} incident neutrons based on PDFs constructed at \SI{0.011}{\eV} and \SI{0.1}{\eV}. 
It can be observed that both the energy distributions are statistically equivalent to the analytical reference curve. 
The results from  our implementation are in excellent agreement with the analytical references.

\begin{figure}[!ht]
    \centering\includegraphics[width=0.45\textwidth]{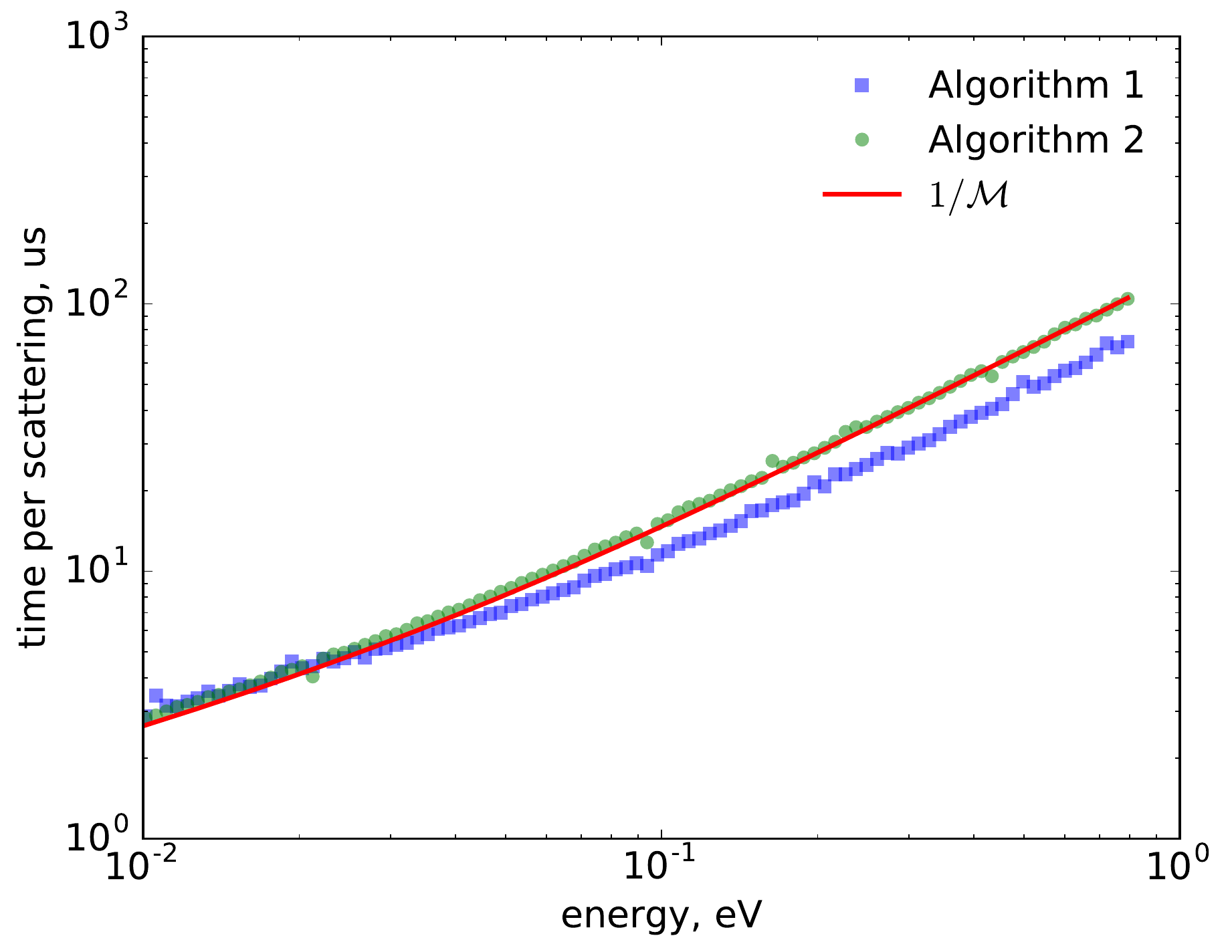}
 \caption{This figure shows the average speed to generate one success final state of a \SI{0.01}{\eV} neutron as a function of the energy where the bounding distributions created. 
 The red curve is normalized to have the same area as the points of the Algorithm~\ref{APredictable}. 
 }
 \label{ffSpeed}
\end{figure}

\replaced{As shown }{Shown} in Fig.~\ref{ffSpeed}, the speed behaviour of Algorithm 1 is well described by $1/\mathcal{M}$. Algorithm 2 shows noticeable speedup when the  energy of the bounding distribution is considerably higher than the incident energy. The speedup grows with incident energy and up to about 45\% when the bounding distribution is constructed at \SI{0.8}{\eV}.

\subsection{Simulation of a water sphere in the free-gas approximation}

We categorise the physics models based on scattering kernels or their derived data as \emph{quantum models}, and the models that sample the scattering as elastic scattering with a thermally-excited nucleus in the centre-of-mass frame as \emph{classical models}. 
When simulating a classical ideal gas system, these two types of models should in principle yield statistically equivalent results, apart from the numerical uncertainties due to the sampling procedure.

To test our numerical implementation, we simulate the volume flux of a free-gas approximated water sphere~\cite{Sublet2009}. 
The  water sphere has a radius of \SI{30}{\cm}, consists of ${^{1}\text{H}_2 ^{16}\text{O}}$, and with a density of \SI{1}{\gram\per\cubic\cm}. 
Neutron scattering with both hydrogen and oxygen are simulated by the kernel driven rejection method introduced in section~\ref{sMethod}. 
The kernel for hydrogen is computed using the free gas model of the LEAPR model in NJOY\added{~\cite{MacFarlane2010}}.
In the simulation, 100 million \SI{1}{\eV} neutrons are initialized with an isotropically distributed random direction at the centre of the sphere. We simulate scattering and capture processes. A neutron is removed from the simulation if it is captured, before reaching the surface of the sphere.

We compare our result with the results predicted by MCNP6~\cite{mcnp6}, in which thermal neutron scattering with hydrogen can be simulated classically, or by two alternative quantum models based on continuous and discrete double differential cross sections, respectively, whereas scattering with oxygen is always modelled using the classical model.
NJOY is used to generate the continuous and discrete cross section files based on~\cite{Brown2018}. The energy grid contains 116 points in the range between \SI{e-5} and \SI{10}{\eV}.

\begin{figure}
\centering
  \centering  \includegraphics[width=0.5\textwidth]{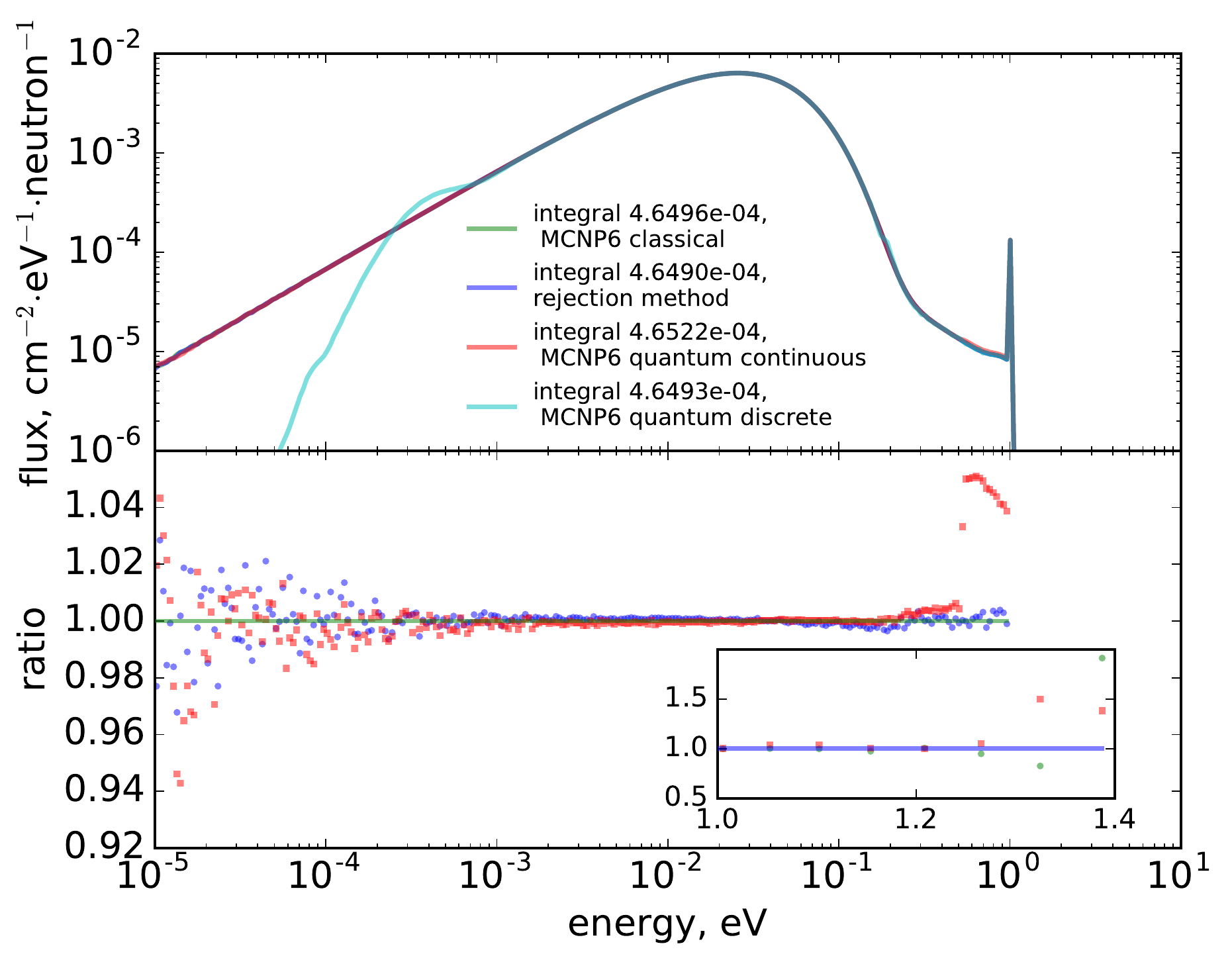}
 \caption{Free-gas treatment of the water sphere. The ratio between the MCNP6 quantum discrete and classical models is not shown. }
 \label{fFree}
\end{figure}

The results are shown Fig.~\ref{fFree}.  
The result from the MCNP6 classical model is used as the reference.
Between \SI{e-3}{\eV} and \SI{1}{\eV}, the fluxes predicted by the rejection method differ only a small fraction of percent from those predicted by the MCNP6 classical model. 
The fluctuations at the low energy end of the rejection method spectrum are due to statistical uncertainties, of which the extent is consistent with that in the spectrum of the MCNP6 continuous model in the same energy region. 
However, between \SI{0.5}{\eV} and the incident energy, i.e.\added{,}~\SI{1}{\eV}, the MCNP6 continuous model \replaced{overestimates }{overestimated} the fluxes by approximately 5\%. 
A later work is performed to use the same continuous cross section with OpenMC~\cite{Romano2015} to try to identify the source of this overestimation. Such overestimation can not be observed from the spectrum given by OpenMC, indicating that this artefact is originated from the sampling method implemented in MCNP6. 
Observed in Fig.~\ref{fFree}, the quantum discrete model in general predicts a realistic integral flux but shows the same artefacts already observed in MCNP~\cite{Sublet2009} below about \SI{1}{\meV}.

\subsection{Applying the rejection method on double differential cross sections}

\begin{figure}
\centering
  \begin{subfigure}[b]{0.45\textwidth}
    \centering\includegraphics[height=0.25\textheight]{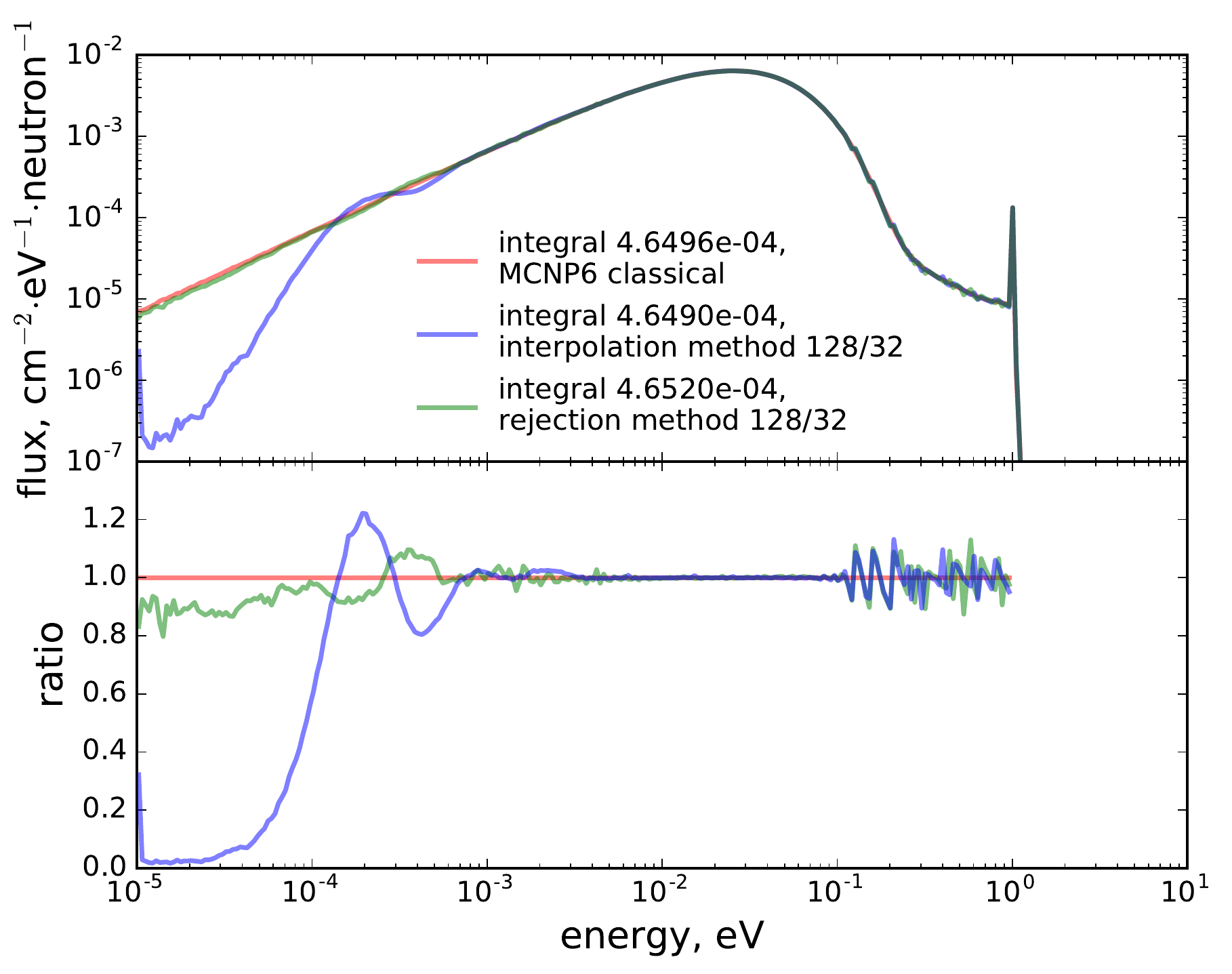}
    \caption{coarse grid ACE file, 128 energy and 32 angular bins}\label{ff128_32_var}
  \end{subfigure}
      ~~~~~~
    \begin{subfigure}[b]{0.45\textwidth}
      \centering\includegraphics[height=0.25\textheight]{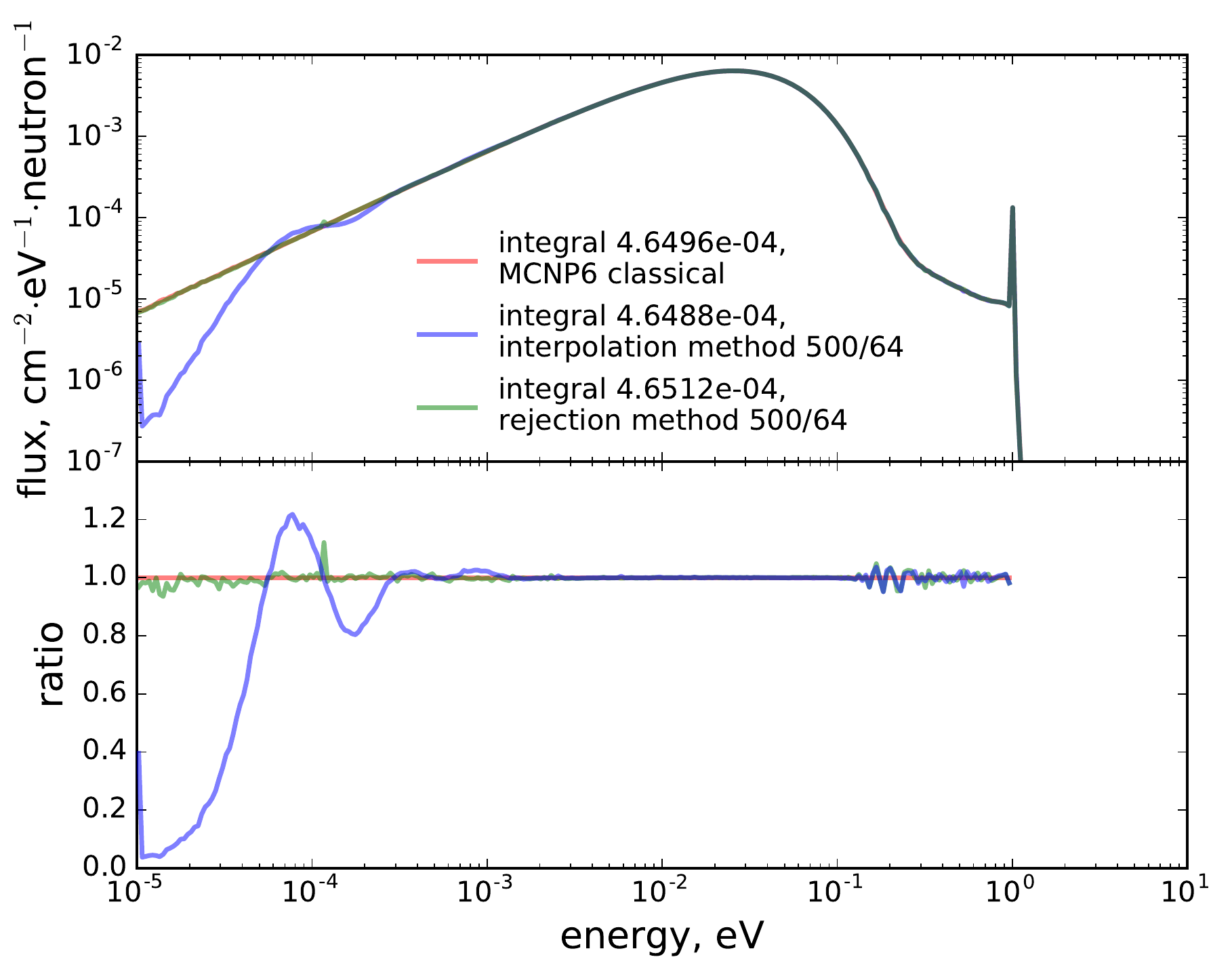}
      \caption{fine grid ACE file,  500 energy and 64 angular bins}\label{ff500_64_var}
    \end{subfigure}
   \caption{Water volume spectra simulation in the free-gas treatment using the discrete double differential cross sections. Spectra are compared with that the reference classical model.}
   \label{fDiscrete500}
\end{figure}

\added[id=R1,remark={R1 C1}]{
For Monte Carlo neutron transport simulations, thermal scattering is conventionally sampled from tabulated double differential cross sections, which are indexed by incident energy, scattered energy and cosine angle. 
The sampling often follows a linear interpolation method, which first generates scattered energy and angle from the distributions of the tabulated energies adjacent to the incident one. The obtained results are interpolated with respect of the incident energy. 
It is possible to apply the rejection method to such data as well.
When applying a dense energy grid in the simulation, the number of sampled neutron final state can be reduced by approximately a factor of two, i.e.\added{,} only the results generated from the distributions of a higher energy $E_i$ need to be considered. }

\replaced[id=R1, remark= R1 C1]{We implemented Algorithm~\ref{ADDXS} in this work for the discrete double differential cross section in the ACE formatted files produced by NJOY. The simulated water sphere spectra are compared with the results obtained by applying the conventional interpolation method to the same data files in Fig.~\ref{ff128_32_var} and~\ref{ff500_64_var} using a coarse and a fine distribution grid, respectively. It can be observed that at}
{The rejection method can be directly applied to discrete cross section in ACE format. Fig.~\ref{fDiscrete500} compares the spectra given by the rejection method and the conventional interpolation method, of which neutron final state is interpolated from the distributions of adjacent energies. At}
the high energy ends, the fluctuations depend highly on the density of the energy and angular grid; while at the low energy ends, the errors are very sensitive to the sampling method chosen. 

\noindent\makebox[\textwidth][c]{%
\begin{minipage}[t]{0.45\textwidth}
  \vspace{0pt}
  \begin{algorithm}[H]
    \label{ADDXS}  
    \caption{for tabulated double differential cross section}
        find smallest i, so that $E \leq E_i$ \\
        \Do{$\mu^\prime \notin\left[-1,1\right]$}{                     
        \Repeat{$-E/k_bT < \beta ^\prime  $}{
             sample an $E^\prime$ from $P(E^\prime\,|\,E_i)$\\
             calculate the $\beta^\prime$ with respect of $E_i$\\
            }
    find j, so that $E^\prime_{j-1} \leq E^\prime \leq E^\prime_j$ \\
    sample $\mu_l$ from $F(\mu \,|\, E^\prime_{j-1},E_i)$\\
    sample $\mu_h$ from $F(\mu \,|\, E^\prime_{j},E_i)$\\
    interpolate a $\mu^\prime$ linearly from  $\mu_l$ and  $\mu_h$\\
        }
        accept $E^\prime$ and $\mu^\prime$  \\
  \end{algorithm}
\end{minipage}%
}

\added[id=R2, remark={R2 C3}]{As for the computational speed, all these four runs are completed between 4.37 and 4.44 CPU hours for \num{e8} source neutrons using the same node of a cluster.  
The rejection method based simulations show a high acceptance rate above 96.9\%, and is only about 1\% faster than the simulations employing the interpolation method, despite the fact that the number of neutron final state sampled in the interpolation method is roughly twice of that in the rejection method. 
As the discrete ACE data structure is highly optimised for sampling speed, the small speed-up factor indicates that a significant part of the CPU time is used by some other parts of the simulations.
}

\subsection{Simulation of a water sphere using a bound hydrogen kernel}

\begin{figure}
\centering
    \centering\includegraphics[width=0.5\textwidth]{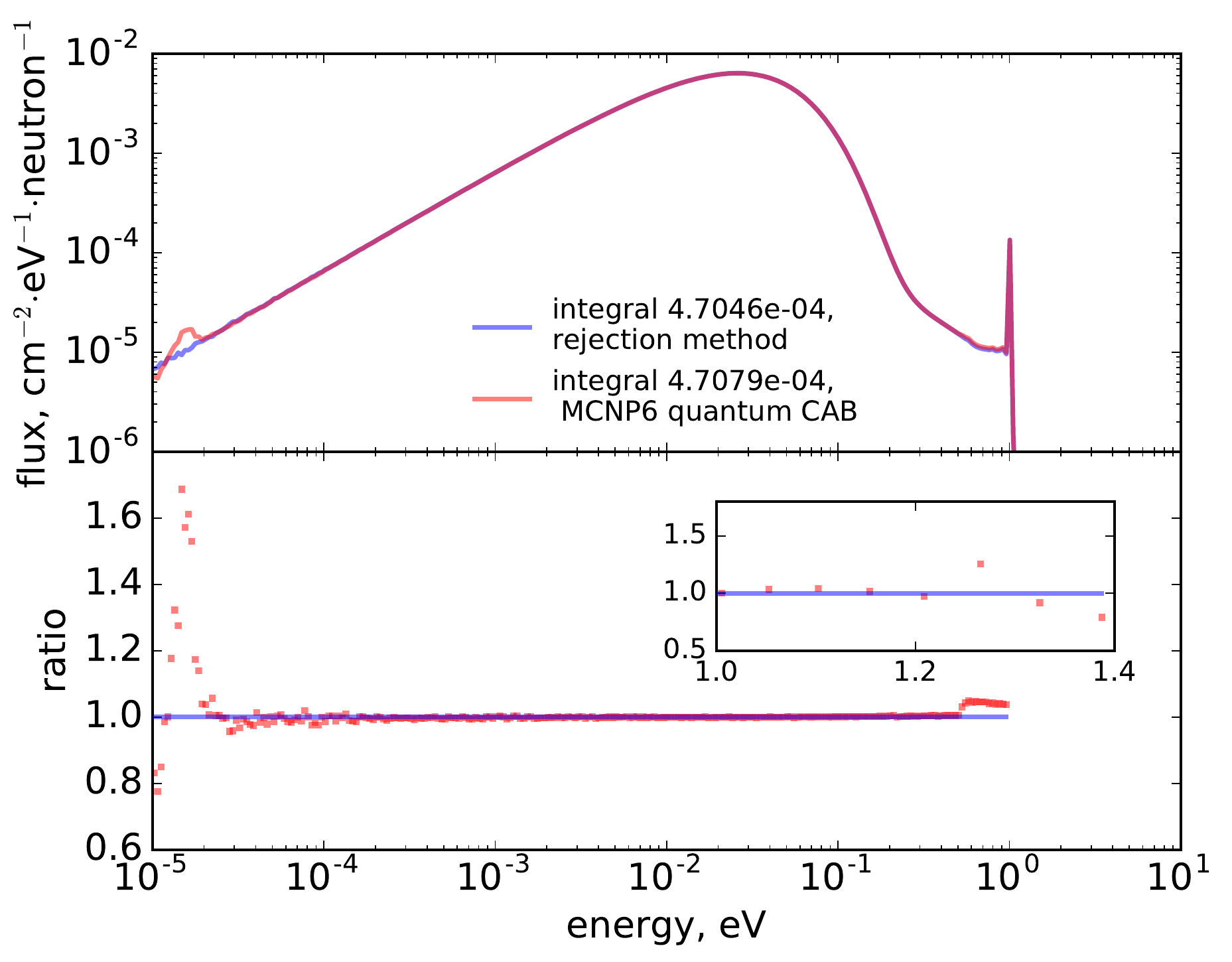}
 \caption{Bound hydrogen treatment of the water sphere. 
 A realistic CAB scattering kernel for hydrogen~\cite{Damian2014} is used.}
 \label{fSphere}
\end{figure}

Replacing the free-gas kernel of hydrogen by one including more realistic binding effects~\cite{Damian2014} \added{that is available in  ENDF/B-VIII.0~\cite{Brown2018} }, we simulate the sphere again. 
\added[id=R2, remark={R2 C2}]{ The $\beta$ grid below 0.5 is refined by a factor of 5, so that the calculated total cross section agree with that of NJOY better than 0.4\% at any incident energy below~\SI{1}{\eV}. }
The resulting volume spectra of our implementation and the MCNP6 quantum continuous model are compared in Fig.~\ref{fSphere}. 
At the high energy end, MCNP6 result shows a similar overestimation as in the free-gas case. At the low energy end, the spectrum from MCNP6 shows structures with fluctuations up to 65\%, corresponding to about 3\% of one standard deviation in that region. As the total cross sections in water below \SI{e-4}{\eV} should be featureless, satisfying the $\sigma\propto1/v$ relation, such \deleted[id=R1, remark=R1 C7]{fluctuates}{fluctuations} are unphysical.

\section{Conclusion}
\label{sConclusion}

The proposed rejection method decouples the dependence on distributions of any lower energy neutrons, therefore requires fewer interpolations.
It is accurate and overcomes some artefacts introduced by conventional interpolation methods. It is also straightforward for an existing code to adapt.  
Our implementation in this paper will be integrated into the software package NCrystal~\cite{Cai2018} in the upcoming releases.

\section*{\added{Acknowledgements}}
\added{
This work was supported in part by the European Union's in the acknowledgements Horizon 2020 research and innovation programme under grant agreement No. 676548 (the BrightnESS project).
Computing resources were provided partly by ESS DMSC.}

\bibliographystyle{elsarticle-num}

\begin{thebibliography}{10}
\expandafter\ifx\csname url\endcsname\relax
  \def\url#1{\texttt{#1}}\fi
\expandafter\ifx\csname urlprefix\endcsname\relax\def\urlprefix{URL }\fi
\expandafter\ifx\csname href\endcsname\relax
  \def\href#1#2{#2} \def\path#1{#1}\fi

\bibitem{hove1954}
L.~{Van Hove}, {Correlations in space and time and Born approximation
  scattering in systems of interacting particles}, Phys. Rev. 95~(1) (1954)
  249--262.

\bibitem{Farhi2009}
E.~Farhi, V.~Hugouvieux, M.~R. Johnson, W.~Kob,
  {{Virtual
  experiments: Combining realistic neutron scattering instrument and sample
  simulations}}, J. Comput. Phys.
   228~(14) (2009) 5251--5261.


\bibitem{MacFarlane2010}
R.~MacFarlane, A.~Kahler,
{{Methods
  for Processing ENDF/B-VII with NJOY}}, Nucl. Data Sheets 111~(12) (2010)
  2739--2890.


\bibitem{bischoff1972}
F.~G. Bischoff, M.~L. Yeater, {Monte Carlo evaluiation of multiple scattering
  and resolution effects in double-differential neutron scattering
  cross-section measurements}, Nucl. Sci. Eng. 48 (1972) 266--280.

\bibitem{Ballinger1999}
C.~T. Ballinger, {The direct S($\alpha$, $\beta$) method for thermal neutron
  scattering}, in: International Conference on Mathematics, Computational
  Methods, and Reactor Physics, 1999, pp. 134--143.

\bibitem{Pavlou2014}
A.~T. Pavlou, W.~Ji,
{{On-the-fly
  sampling of temperature-dependent thermal neutron scattering data for Monte
  Carlo simulations}}, Ann. Nucl. Energy
  71 (2014) 411--426.

\bibitem{hart2017}
S.~W. Hart, G.~I. Maldonado,
{{Implementation
  of the direct S($\alpha$, $\beta$) method in the KENO Monte Carlo code}},
  Ann. Nucl. Energy 101 (2017) 270--277.


\bibitem{Everett1983}
C.~J. Everett, E.~D. Cashwell, {A Third Monte Carlo Sampler}, Tech. Rep.
  LA-9721-MS, Los Alamos National Laboratory (March 1983).

\bibitem{trkov2012endf}
A.~Trkov, M.~Herman, D.~Brown, et al., {ENDF}-6 formats manual, Data Formats
  and Procedures for the Evaluated Nuclear Data Files ENDF/B-VI and ENDF/B-VII,
  National Nuclear Data Center Brookhaven National Laboratory, Upton, NY (2012)
  11973--5000.

\bibitem{Damian2014}
J.~{M{\'{a}}rquez Dami{\'{a}}n}, J.~Granada, D.~Malaspina,
{{CAB models
  for water: A new evaluation of the thermal neutron scattering laws for light
  and heavy water in ENDF-6 format}}, Ann. Nucl. Energy 65 (2014)
  280--289.


\bibitem{squiresnew}
G.~L. Squires, {Introduction to the Theory of Thermal Neutron Scattering},
  Cambridge, 1997, Ch. 2.3.

\bibitem{Weston2007}
W.~M. Stacey, Nuclear Reactor Physics, 2nd Edition, Wiley-VCH, Ch. 12.2.

\bibitem{Brown2018}
D.~A. Brown, et al.,
 {ENDF/B-VIII.0:
  The 8 th Major Release of the Nuclear Reaction Data Library with
  CIELO-project Cross Sections, New Standards and Thermal Scattering Data},
  Nucl. Data Sheets 148 (2018) 1--142.

\bibitem{Sublet2009}
J.-C. Sublet, D.~Cullen, R.~MacFarlane, {How accurately can we calculate fast
  neutrons slowing down in water?}, Nucl. Technol.
  168~(2) (2009) 293--297.

\bibitem{mcnp6}
J.~Goorley, et~al., {Initial MCNP6 Release Overview - MCNP6 version 1.0}, Tech.
  Rep. LA-UR-13-22934, {Los Alamos National Laboratory} (2013).

\bibitem{Romano2015}
P.~K. Romano, N.~E. Horelik, B.~R. Herman, A.~G. Nelson, B.~Forget, K.~Smith,
{{OpenMC:
  A state-of-the-art Monte Carlo code for research and development}}, Ann. Nucl. Energy 82 (2015) 90--97.


\bibitem{Cai2018}
X.~X. Cai, T.~Kittelmann, N{C}rystal, https://doi.org/10.5281/zenodo.853186.

\end{thebibliography}

\end{document}